\documentclass[12pt]{article}

\usepackage{epsfig}

\title{
The Role of Phonon Mechanism in Electron Coupling
\author{Ludmila Joukovskaya\\
~\\
{\it Physics Department, Moscow State University}\\
{\it Vorobievi Gori, 119899 Moscow, Russia}\\
{\tt l\_joukovskaya@mtu-net.ru}} }

\begin{document}

\date{}
\maketitle

\begin{abstract}
In this article in the framework of generalized Frolich model we
consider electron-spin-phonon system. A parameter of
electron-spin-phonon interaction is found. It is shown its
interrelation with experiments on isotope-effect on high
temperature superconductors and asymptotic transformation to BCS
theory \cite{BS,BCS}.
\end{abstract}



$$
$$

Experimental study of high temperature superconductors have shown
that electron-phonon interaction decreases with the raise of
critical temperature $T_c$ \cite{CC}. Thus electron coupling based
only on phonon mechanism -- the technique used to explain low
temperature superconductivity -- fails to explain high critical
temperatures \cite{MAM}. One can propose that in order to explain
high $T_c$ a theory with more complex mechanism in which the role
of phonons decreases but does not vanish should be considered.

In this article in the framework of generalized Frolich model we
will consider electron-spin-phonon interaction which is formed by
the sum of electron-phonon, electron-spin and electron-spin-phonon
interacting terms.

We start from the Hamiltonian of electron-spin-phonon system
\cite{SS}
\begin{eqnarray}
\label{hamil}
 {\hat{H}}_{e+ph+s}=\hat{H}_{e}+\hat{H}_{ph}+\hat{H}_{s}+\hat{H}_{e-ph}
 +\hat{H}_{e-s}+\hat{H}_{s-ph}+\hat{H}_{e-s-ph}
\end{eqnarray}
Now let us write the terms corresponding to the interaction of
electrons with phonons and longitudinal  spin modes
\begin{eqnarray}
\label{eq2} {\hat{H}_{e-ph}}=\frac{1}{\sqrt{V}} \sum
\frac{g_{ph}}{\omega_D}i
\sqrt{\frac{{\hbar}{\omega}_{ck{\nu}}}{2}}({n_{\nu
}}{e_{\nu}})(\hat{b}_{k_1 \nu }+\hat{b}_{-k_1 \nu}^{+})
\hat{c}^{+}_{k_2 }\hat{c}_{(k_2-k_1)}
\end{eqnarray}
\begin{eqnarray}
\label{eq3}
 {\hat{H}_{e-s}}=\eta \frac{(p_{\nu e}n_{\nu })}{\sqrt{\mu _e}}
 \frac{1}{\sqrt{V}}\sum i\sqrt{\frac{{\hbar}{\omega}_{sk}}{2}}
 (\hat{a}_{k_1 z}-\hat{a}^+_{-k_1 z})\hat{c}^{+}_{k_2}\hat{c}_{(k_2-k_1)}
\end{eqnarray}
\begin{eqnarray}
 \label{eq4}
 {\hat{H}_{e-s-ph}}=\zeta \eta
 \frac{(p_{\nu^{\prime}e}n_{\nu^{\prime}})}{\sqrt{\mu_e}}(n_\nu e_3) \frac{1}{\sqrt{V}}
 \sum i\sqrt{\frac{{\hbar}{\omega}_{ck3}}{2}}
 (\hat{b}_{k_1 3}-\hat{b}_{-k_1 3 }^+)\hat{c}^{+}_{k_2
}\hat{c}_{(k_2-k_1)}
\end{eqnarray}
here $g_{ph}$, $\zeta$, $\eta$ -- are the coupling constants of
electron and phonon subsystem, spin and phonon subsystem, electron
and spin subsystem,
 $a^{+}$, $a$;~~$b^{+}$, $b$;~~$c^{+}$, $c$ --
creation and annihilation operators of magnons, phonons, and
electrons, $\omega_D$ -- is the Debye   frequency, that is the
maximum frequency of phonons which take part in the interaction,
vectors $e_\nu$ represent phonon polarizations ($\nu=1,2,3$),
$n_\nu=k_\nu/|k_\nu|$, where $k_\nu$ is the wave vector, $c$ --
the sound velocity, $\omega_{sk}={J_0 s}\sqrt{(k/k_c)^2-1}$ -- the
frequency of the longitudinal  spin mode linearly coupled with
phonons, $J_0$ -- the exchange potential, $s=1/2$ is the spin of
electron, $k_c=2\pi/r_c$, $r_c$ -- the exchange correlation
radius, ${\omega}_{ck1,2}=ck$ and
$\omega_{ck3}=ck\sqrt{1+\zeta^{2}}$ -- the frequency of the phonon
mode linearly coupled with the longitudinal spin mode, $p_{e\nu}$
-- momentum of electron, $\mu_e$ -- electron mass.

Using generalized Bogoliubov's transformation we represent
operators $a^{+}$, $a$ and $b^+$, $b$ in the following form
~\cite{6}
\begin{equation}
\hat{a}_{zk}=u_{zzk}\hat{c}_{zk}+v^{*}_{zz;-k}\hat{c}^{+}_{z;-k}
+u_{z3k}\hat{d}_{3k}+v^{*}_{z3;-k}\hat{d}^{+}_{3;-k}
\end{equation}
\begin{equation}
\hat{a}_{z;-k}^{+}=u^{*}_{zz;-k}\hat{c}^{+}_{z;-k}
+v_{zzk}\hat{c}_{zk}
+u^{*}_{z3;-k}\hat{d}^{+}_{-k;3}+v_{z3k}\hat{d}_{k;3}
\end{equation}
\begin{equation}
\hat{b}_{3k}=u_{33k}\hat{d}_{3k}+v^{*}_{33;-k}\hat{d}^{+}_{3;-k}
+u_{3zk}\hat{c}_{zk} +v^{*}_{3z;-k}\hat{c}^{+}_{z;-k}
\end{equation}
\begin{equation}
\hat{b}_{3;-k}^{+}=u^{*}_{33;-k}\hat{d}^{+}_{3;-k}
+v_{33k}\hat{d}_{3k}
+u^{*}_{3z;-k}\hat{c}^{+}_{z;-k}+v_{3zk}\hat{c}_{3k}
\end{equation}

The integral equation for the energy gap is given by
$$
\Delta(\varepsilon)=\int_{0}^{\hbar \omega_c}
\frac{\Delta(\varepsilon')}{\varepsilon'}
Q(\varepsilon,\varepsilon')
th\frac{\varepsilon'}{2T}d\varepsilon',
$$ where $\omega_c$ is close to the order of maximum coupled
spin-phonon oscillations. The kernel of this integral equation can
be written in the form   $Q=Q_{e-ph}+Q_{e-s}+Q_{e-s-ph}$. Each
term can be presented as the sum of the Green functions using
coefficients in (\ref{eq2})-(\ref{eq4}) ~\cite{SBS}. Considering
the long-waves limit we get
$$
Q_{e-ph}(k)=N(0) \frac{2g^{2}_{ph}}{3\omega_{D_{0}}}+
N(0)
\frac{g^{2}_{ph}}{3\omega_{D_{0}}\sqrt{1+\zeta^2}}[|u_{33k}+v_{33k}|^{2}\frac{\omega
_{ck3} }{\varepsilon_{3k}}+|u_{3zk}+v_{3zk}|^{2}\frac{\omega
_{ck3} }{\varepsilon_{sk}}]
$$
\begin{eqnarray}
\label{eq6} Q_{e-s}(k)=N(0)
 \frac{p_{F}^{2}}{3 \mu _{e}}\eta^{2}
[|u_{zzk}-v_{zzk}|^{2}\frac{\omega _{sk}
}{\varepsilon_{sk}}+|u_{z3k}-v_{z3k}|^{2}\frac{\omega _{sk}
}{\varepsilon_{3k}}]
\end{eqnarray}
$$
Q_{e-s-ph}(k)=N(0)\zeta^{2}
 \frac{p_{F}^{2}}{9 \mu_e}\eta^{2}
[|u_{33k}-v_{33k}|^{2}\frac{\omega _{ck3}
}{\varepsilon_{3k}}+|u_{3zk}-v_{3zk}|^{2}\frac{\omega _{ck3}
}{\varepsilon_{sk}}]
$$
here $\omega_{D_{0}}$ -- is the representative value of Debye
frequency in the BCS theory of low temperature superconductors,
while $\omega_{D_{1}}=\omega_{D_{0}}\sqrt{1+\zeta^2}$ is the Debye
frequency in the electron-spin-phonon system, $\varepsilon_{sk}$
and $\varepsilon_{3k}$ -- are the frequencies of coupled
spin-phonon oscillations, and $N(0)$ -- is the density of electron
states on Fermi surface. Please note that $g_{ph}$ is equal to
zero when $k>2k_F$, while $\zeta$ and $\eta$ are constants.

Let us now consider the additional contribution to the Coulomb
repulsion of electrons which appears as a result of spin-electron
interaction. In the simplest case of isotropic Fermi surface this
contribution is given by \cite{Kit}
\begin{eqnarray}
\label{eq7} \Delta \mu = N(0)\frac{p^{2}_{F}}{3 \mu_{e}}\eta^{2}
\end{eqnarray}
Now let us write the value of $Q_{e-s}(k)$ with this contribution
\begin{equation}
\label{Q2}
Q_{e-s}(k)=N(0) \frac{p_{F}^{2}}{3 \mu _{e}}\eta^{2}
[|u_{zzk}-v_{zzk}|^{2}\frac{\omega _{sk}
}{\varepsilon_{sk}}+|u_{z3k}-v_{z3k}|^{2}\frac{\omega _{sk}
}{\varepsilon_{3k}}-1]
\end{equation}
Substituting the explicit values of functions $u_{33k}$,
$v_{33k}$, $u_{3zk}$, $v_{3zk}$, $u_{z3k}$, $v_{z3k}$, $u_{zzk}$,
$v_{zzk}$, to (\ref{eq6}), (\ref{Q2}) we obtain
$$
Q_{e-ph}(k) =N(0) \frac{2g^{2}_{ph}}{3\omega_{D_{0}}} + \nu
(\varepsilon _{F})
\frac{g^{2}_{ph}}{3\omega_{D_{0}}\sqrt{1+\zeta^2}}q_{e-ph}(k)
$$
\begin{eqnarray}
\label{eq28} Q_{e-s}(k)= N(0) \frac{p_{F}^{2}}{3 \mu _{e}}\eta^{2}
\left[ q_{e-s} -1\right]
\end{eqnarray}
$$
Q_{e-s-ph}(k)=N(0) \frac{p_{F}^{2}}{9 \mu _{e}}\eta^{2}\zeta^{2}
q_{e-s-ph}(k)
$$
where
$$
q_{e-ph}(k)=\frac{z^{2}\omega ^{2}_{sk}\omega
^{2}_{ck3}}{(\varepsilon ^{2}_{3k}- \varepsilon
^{2}_{sk})(\varepsilon ^{2}_{3k}-\omega ^{2}_{ck3})}  +
\frac{(\varepsilon ^{2}_{sk}-\omega ^{2}_{sk})} {(\varepsilon
^{2}_{sk}-\varepsilon ^{2}_{3k})}
$$
$$
q_{e-s}(k)=\frac{z^{2}\omega ^{4}_{sk}\omega ^{2}_{ck3}}{
\varepsilon ^{2}_{sk}(\varepsilon ^{2}_{sk}- \varepsilon
^{2}_{3k})(\varepsilon ^{2}_{sk}-\omega ^{2}_{sk})}+
\frac{\omega^{2}_{sk} (\varepsilon ^{2}_{3k}-\omega ^{2}_{ck3}) }
{ \varepsilon ^{2}_{3k} (\varepsilon ^{2}_{3k}-\varepsilon
^{2}_{sk})}
$$
$$
q_{e-s-ph}(k)=
 \frac{z^{2}\omega ^{2}_{sk}\omega
^{4}_{ck3}}{ \varepsilon ^{2}_{3k}(\varepsilon ^{2}_{3k}-
\varepsilon ^{2}_{sk})(\varepsilon ^{2}_{3k}-\omega ^{2}_{3k})}+
\frac{ \omega ^{2}_{3k} (\varepsilon ^{2}_{sk}-\omega ^{2}_{sk}) }
{ \varepsilon ^{2}_{sk} (\varepsilon ^{2}_{sk}-\varepsilon
^{2}_{3k})}
$$

Let us now compute $Q_{e-ph}(k)$, $Q_{e-s}(k)$, and
$Q_{e-s-ph}(k)$ explicitly. We introduce the following functions
\begin{eqnarray}
\label{eq101}
 w_{e-ph}=q_{e-ph}(k);
\end{eqnarray}
$$
w_{e-s}=\frac{1}{1+\zeta^2}q_{e-s}(k);
$$
$$
w_{e-s-ph}=\frac{1}{1+\zeta^2}q_{e-s-ph}(k).
$$

\begin{figure}
\begin{center}
\epsfig{file=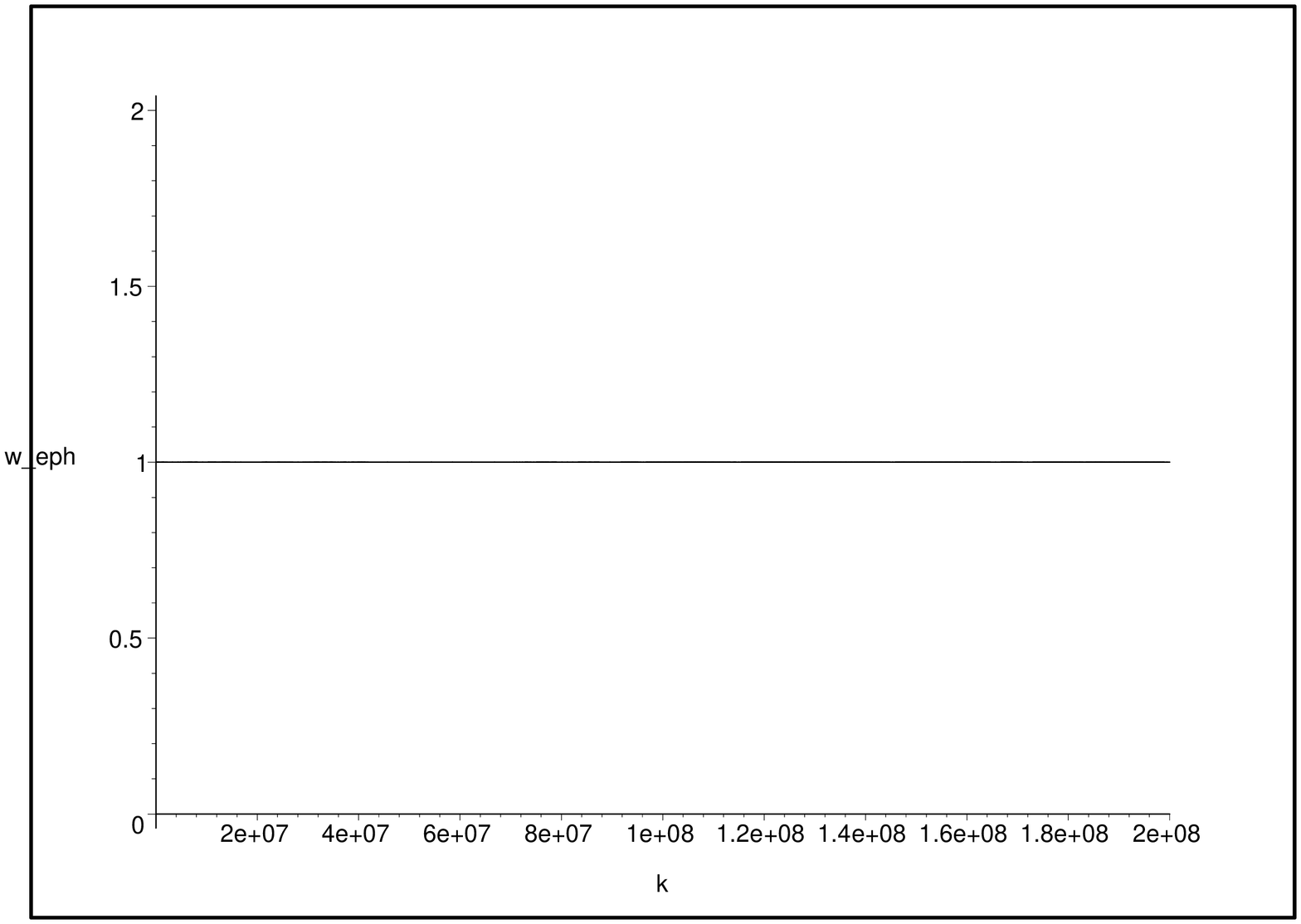,width=3.5cm,angle=-90}
\epsfig{file=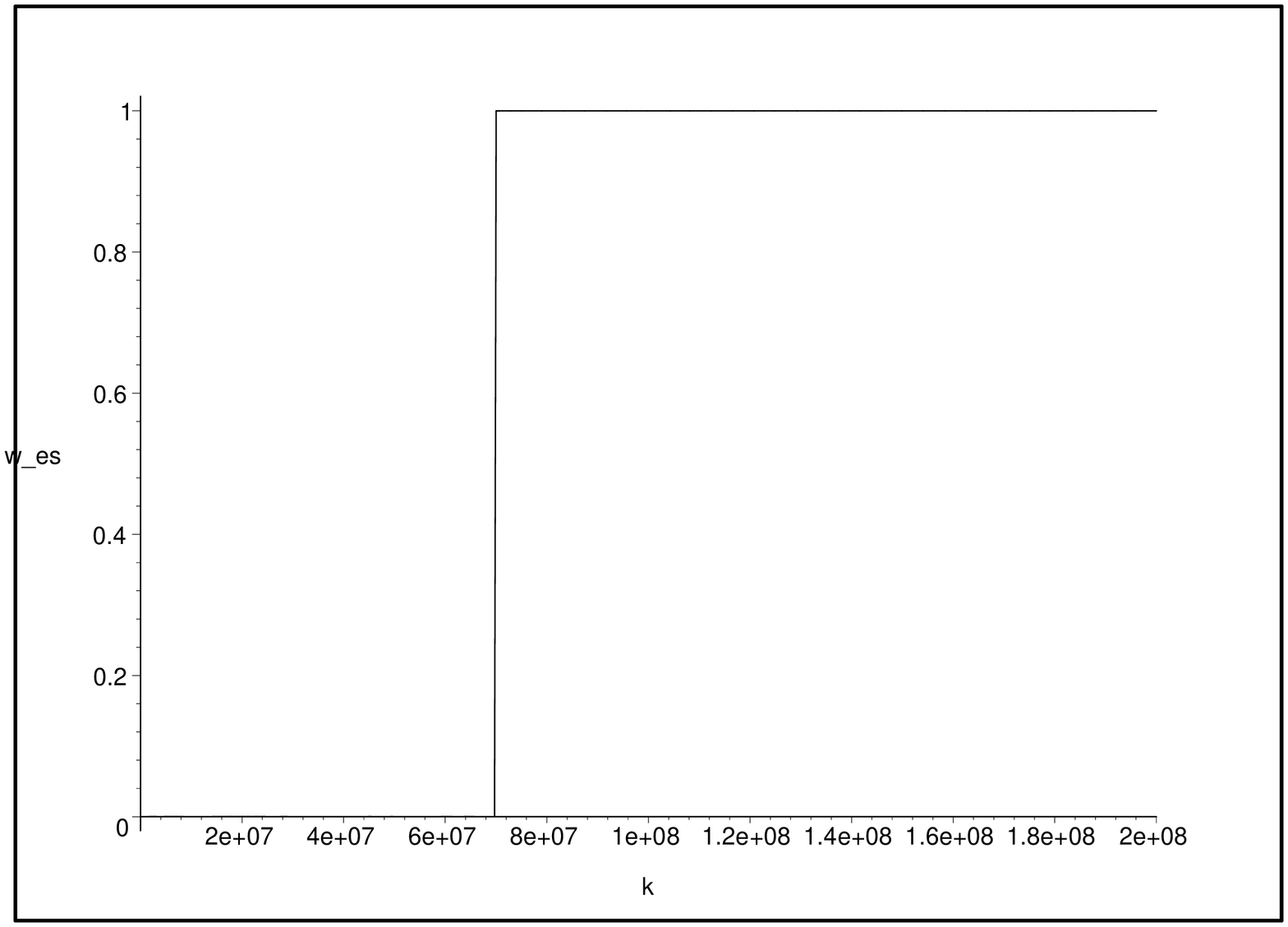,width=3.5cm,angle=-90}
\epsfig{file=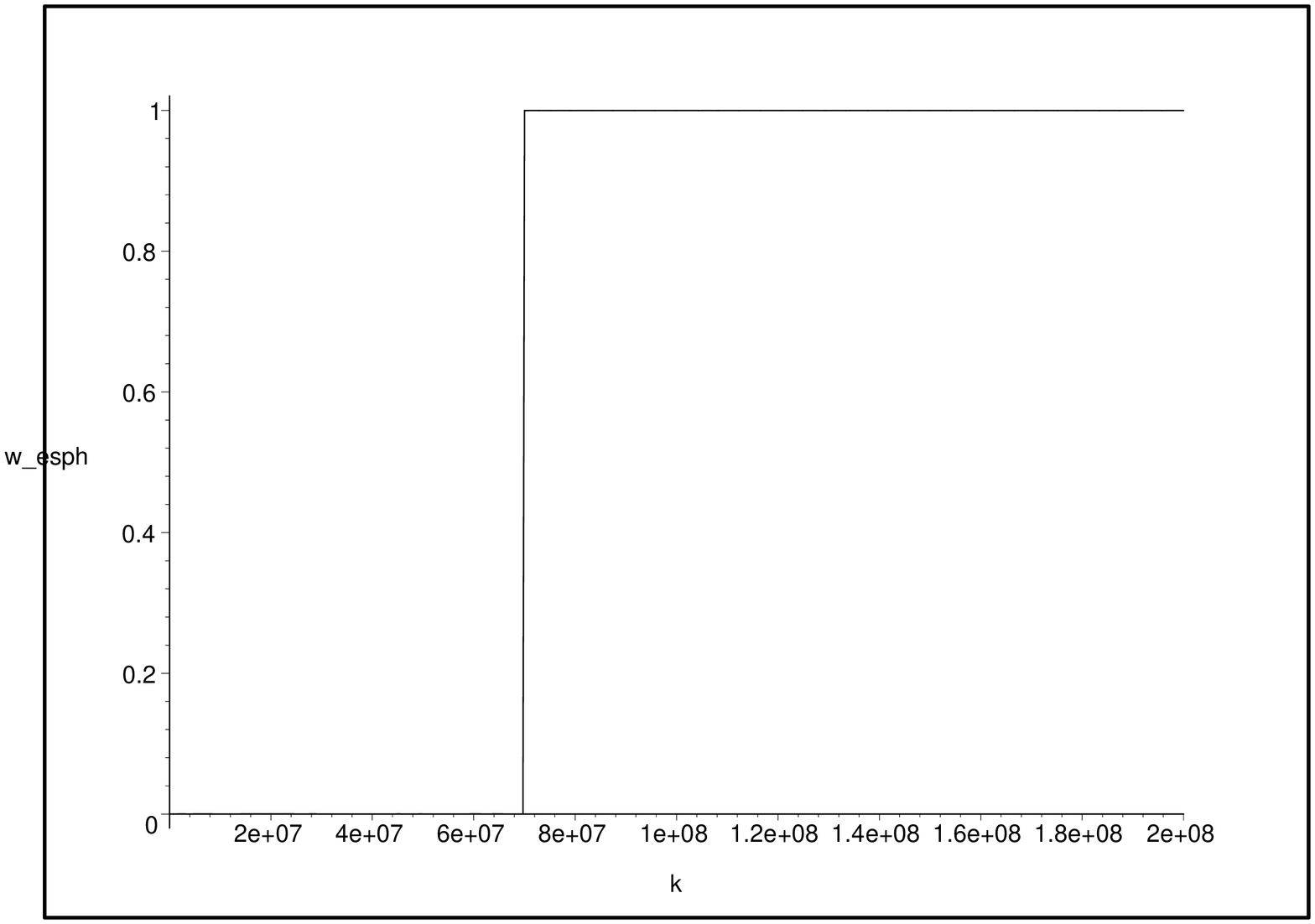,width=3.5cm,angle=-90} \caption{ The results
of numerical simulation of the functions $w_{e-ph},w_{e-s}$ and $
w_{e-s-ph}$ (\ref{eq101}). These plots illustrates the equalities
(\ref{EephEtAl}). The upper left is the plot of $w_{e-ph}$ as a
function of $k$, the upper right shows the picture of $w_{e-s}$,
and the lower -- $ w_{e-s-ph}$. Let us note that for the values of
$\zeta$ in the range $0<\zeta<10$ all functions (\ref{eq101}) have
the same form. One can see that for $k>k_c$, i.e. in the area
where the coupling is performed the functions $w_{e-ph}$,
$w_{e-s}$, and $ w_{e-s-ph}$ are equal to $1$.}
\end{center}
\end{figure}

For $k>k_c$ the functions $w_{e-ph}$, $w_{e-s}$, and $w_{e-s-ph}$
are equal to $1$. Let us note that since the pairing is performed
in the range $r<r_c$ and $k_c=2\pi/r_c$ the range $k>k_c$ is of
our interest. Thus using (\ref{eq101}) we get
\begin{equation}
\label{EephEtAl} q_{e-ph}(k)= 1;
\end{equation}
$$
q_{e-s}(k)=1+\zeta^2;
$$
$$
q_{e-s-ph}(k)=1+\zeta^2.
$$
Results of numerical simulation of these functions are presented
on Fig.1.

Using these results we simplify (\ref{eq28}) getting
\begin{eqnarray}
Q_{e-ph}(k)= N(0) \frac{2g^{2}_{ph}}{3\omega_{D_{0}}}+N(0)
\frac{g^{2}_{ph}}{3\omega_{D_{0}}\sqrt{1+\zeta^2}}
\end{eqnarray}
$$
Q_{e-s}(k)=N(0)
 \frac{p_{F}^{2}}{3 \mu _{e}}\eta^{2}
\zeta^2
$$
$$
Q_{e-s-ph}(k)=N(0)\zeta^{2}
 \frac{p_{F}^{2}}{9 \mu _{e}}\eta^{2}
(1+\zeta^2).
$$

Now $Q(k)$ will be
\begin{eqnarray}
\label{rez} Q(k)= N(0) \left[ \frac{2g^{2}_{ph}}{3\omega_{D_0}}+
\frac{g^{2}_{ph}}{3\omega_{D_0}\sqrt{1+\zeta^2}}+
\frac{p_{F}^{2}}{9 \mu_{e}}{\zeta}^{2}\eta^{2}(4+{\zeta}^{2})
\right]
\end{eqnarray}

Let us consider two frequency ranges. The range $\omega <
\omega_{D_0}$ corresponds to low temperature super conductivity,
while the range $\omega>\omega_{D_0}$ corresponds to high
temperature superconductivity.

In the range $\omega>\omega_{D_0}$ the expression (\ref{rez})
takes the form
\begin{equation}
\label{Q1}
Q(k)= N(0) \left[
\frac{g^{2}_{ph}}{3\omega_{D_0}\sqrt{1+\zeta^2}}+
\frac{p_{F}^{2}}{9 \mu_{e}}{\zeta}^{2}\eta^{2}(4+{\zeta}^{2})
\right]
\end{equation}
From (\ref{Q1}) we see that electron-phonon interaction decreases
with the raise of critical temperature (that is with the raise of
frequency). This fact is supported by experiments on
isotope-effect, which shows that electron-phonon interaction
should be taken into consideration ~\cite{9}. Although since
electron-phonon interaction alone is not enough to explain high
critical temperatures, one should consider more complex
interaction, like spin-electron-phonon interaction, as it is done
in the present work.It was shown in experiments on La$_2$CuO$_4$
that for $T>T_N$ spin fluctuations have high energy and the
velocity of spin fluctuations ($\sim10^6$ cm/s) is greater than
the sound velocity ($\sim10^5$ cm/s) ~\cite{10}.

In the range $\omega<\omega_{D_0}$ we have
$$
\zeta=\frac{\sqrt{3} g \hbar k_{c}}{\sqrt{J_{0}s M}}\rightarrow 0
$$
here $M$ is the reduced mass of ion, $g=U/J_0$, where $U$ is the
electron-ion potential. Now the expression (\ref{rez}) takes the
form
$$
Q(k)= N(0) \frac{g^{2}_{ph}}{\omega_{D_0}},
$$
which is exactly the electron-phonon interaction term in BCS
theory.

Let us note that spin fluctuations exist in all superconductors.
Although in low temperature superconductors this fluctuations are
small and thus do not seriously affect electron
coupling\cite{11,12,13}. On the other hand in high temperature
superconductors these fluctuations increases due to the second
term in (\ref{Q1}).

\vspace{0.5cm}
{\large Acknowledgment}

The author is grateful to B.I. Sadovnikov for constant attention
to this work, the author is also grateful to M.A. Savchenko for
fruitful discussions.

\end{document}